\documentclass[sigconf,screen,nonacm]{acmart}
\setcopyright{none}
\keywords{Conditional memory, lookahead scheduling, CXL, tiered memory, heterogeneous inference}
\usepackage{siunitx}
\usepackage{pifont}
\setcitestyle{numbers,square}
\ifdefined\noindentparagraph
  \newcommand{\runinheading}[1]{\noindentparagraph{\normalfont\bfseries #1.}}
\else
  \newcommand{\runinheading}[1]{\paragraph*{\textbf{#1}}}
\fi
\makeatletter
\@ifundefined{hyxmp@parse@acmart}{}{%
  \let\hyxmp@parse@acmart\relax
  \hypersetup{keeppdfinfo,pdfcontactemail={haojiazhang@nudt.edu.cn}}}
\makeatother

\begin{document}
\title{Scaling Engram Inference with Tiered CXL Memory and Cross-Block Execution}

\author{Haojia Zhang}
\authornote{Corresponding author. Email: \href{mailto:haojiazhang@nudt.edu.cn}{haojiazhang@nudt.edu.cn}.}
\affiliation{\institution{National University of Defense Technology}\city{Changsha}\country{China}}
\author{Hanzhi Xun}
\affiliation{\institution{National University of Defense Technology}\city{Changsha}\country{China}}
\author{Yang Ou}
\affiliation{\institution{National University of Defense Technology}\city{Changsha}\country{China}}
\author{Jiaqing Xu}
\affiliation{\institution{National University of Defense Technology}\city{Changsha}\country{China}}
\author{Fangxu Lv}
\affiliation{\institution{National University of Defense Technology}\city{Changsha}\country{China}}
\author{Zicong Wang}
\affiliation{\institution{National University of Defense Technology}\city{Changsha}\country{China}}
\author{Wei Chen}
\affiliation{\institution{National University of Defense Technology}\city{Changsha}\country{China}}
\author{Lizhou Wu}
\affiliation{\institution{National University of Defense Technology}\city{Changsha}\country{China}}
\author{Mingche Lai}
\affiliation{\institution{National University of Defense Technology}\city{Changsha}\country{China}}
\renewcommand{\shortauthors}{Zhang et al.}

\begin{abstract}
Engram replaces part of knowledge retrieval in large language models with sparse lookup, reserving more Transformer blocks for compositional reasoning and increasing effective computational depth. However, scaling its embeddings poses two challenges. First, placement of embeddings must balance capacity, cost, and access latency. Because Embedding access windows differ across Engram blocks, uniform placement cannot readily meet their latency requirements; shared access contention and entry transfers can add further delay. Second, hidden-state dependencies place Gate, Conv, and their transfers on the critical path, while earlier execution must balance sufficient overlap against hidden-state differences as workloads change. We present FlexEngram, a decode-focused framework with three mechanisms. First, block-wise disaggregated placement uses HBM for backbone state, host DRAM for Embedding outputs, CXL-DRAM for early Engram blocks' embeddings, and CXL-PCM for the rest. Sharing, batched reads, and reusable pinned buffers reduce delivery costs. Second, dependency-aware heterogeneous execution uses earlier hidden states to overlap Gate and Conv with the backbone through CPU-accelerated, GPU-accelerated, or ASIC-accelerated. At the target, the backbone adds their output delta to the backbone hidden state; CUDA fusion reduces residual GPU overhead. Third, an adaptive minimum lookahead distance scheduler predicts the minimum CPU-accelerated lookahead distance from Gate--Conv and transfer costs and observed block times. GPU-accelerated and ASIC-accelerated use one-block lookahead. On Qwen3-4B and Llama-3.1-8B-Instruct, FlexEngram improves average decode throughput by 3.15\% over VanillaEngram and 3.23\% over PoolingEngram. The scheduler achieves a 100\% match between predicted and measured minimum lookahead distances across multiple workloads. Eight Engrams outperform two-module VanillaEngram in throughput. Across the three paths, each added Engram incurs 0.24\% mean throughput loss, an 83.46\% smaller decline than VanillaEngram, supporting larger conditional memory with minimal overhead.
\end{abstract}

\maketitle

\section{Introduction}\label{sec:I}

Large language models (LLMs) rely on both knowledge retrieval and compositional reasoning. Mainstream LLM frameworks use attention and feed-forward networks (FFNs)~\cite{NIPS2017_3f5ee243} to perform both tasks. Early Transformer blocks (hereafter blocks; Engram inserted at Transformer block $i$ is called Engram block $i$) perform part of the knowledge retrieval computation. To reduce this computation, sparse key/value memory layers store learned knowledge and directly read the memory contents selected for each token \cite{NEURIPS2019_9d8df73a}. Specifically, Engram implements this knowledge retrieval approach with learned Engram embedding tables, hereafter embeddings, that store memory contents for sparse lookup \cite{cheng2026conditionalmemoryscalablelookup}. For each token, Engram directly reads the selected memory contents from embeddings, then uses Linear, Gate, and Conv to produce a delta for addition to the backbone hidden state before the target block's attention and FFN computation. By replacing part of this knowledge retrieval computation with sparse lookup, Engram reserves more blocks for compositional reasoning and increases effective computational depth.

Prior work improves conditional-memory inference through memory organization, data access, and operator optimization. VanillaEngram~\cite{cheng2026conditionalmemoryscalablelookup} offloads embeddings to host DRAM and overlaps Embedding accesses determined by token IDs with Transformer execution; PoolingEngram~\cite{10.1145/3805621.3807654} extends capacity through shared CXL memory and efficient fine-grained reads. For flash-backed memory, PageGram~\cite{lehmann2026pagegram} groups heads of the same n-gram order within a page to reduce random flash reads. TF-Engram~\cite{ma2026tfengramtrainfreeengramssdbacked} organizes phrase memory across GPU, DRAM, and SSD and predicts next-token candidates for prefetching. At the operator level, KernelEngram~\cite{du2026kernelsmithunifiedrecipeevolutionary} uses evolutionary kernel optimization to fuse Gate, normalization, Conv, and residual updates on the GPU.

Large-scale Engram inference faces two challenges. First, embeddings placement must balance capacity, cost, and access latency. Low-latency memory is costly, whereas denser, higher-latency media can delay Embedding. Access windows vary by Engram block and are shortest near the beginning, so uniform placement cannot readily meet position-dependent latency needs. Shared contention and entry transfers can further delay delivery. Second, Gate and Conv depend on hidden states. Waiting for the target state places their execution and transfers on the critical path; earlier states enable overlap but change Gate's input and may affect output. Gate--Conv, transfer, and block times also vary with workload and execution path. Fixed lookahead can therefore provide insufficient overlap or use an unnecessarily distant hidden state.

FlexEngram coordinates placement of embeddings by Engram block with early execution. Backbone weights, key and value caches, and working states remain in GPU high-bandwidth memory (HBM); host DRAM supplies reusable staging buffers. Early Engram blocks' embeddings use low-latency CXL-DRAM; the remaining embeddings use denser CXL-PCM; CXL sharing lets instances reuse static embeddings without replicas. Batched Embedding reads and contiguous pinned buffers reduce entry delivery costs. Once Hash, Embedding, Concat, and Linear have produced key and value, an earlier hidden state launches Gate--Conv alongside subsequent Transformer computation. When the backbone reaches the target block, it adds the delta produced by early Gate--Conv execution to the backbone hidden state, before that block's attention and FFN computation. To support this early Gate--Conv execution, FlexEngram provides three paths: CPU-accelerated with fused CPU kernels, GPU-accelerated with an independent GPU stream, and ASIC-accelerated with a dedicated Engram accelerator. FlexScheduler predicts the minimum CPU-accelerated lookahead distance from Gate--Conv and transfer costs and completed-block timings, whereas GPU-accelerated and ASIC-accelerated run one block ahead. Across Qwen3-4B and Llama-3.1-8B-Instruct, these paths improve average decode throughput by 3.15\% over VanillaEngram and 3.23\% over PoolingEngram. In Qwen's scaling experiment, average throughput loss per added Engram is 0.24\%, 83.46\% below VanillaEngram's decline. To our knowledge, this is the first system combining Engram block-wise CXL-DRAM/CXL-PCM tiering by Embedding access window with workload-adaptive CPU-accelerated lookahead.

\begin{figure}[t!]
{\centering
\includegraphics[width=\linewidth,alt={Hash maps token IDs to indices for Embedding; Concat joins the read vectors, and Linear produces key and value vectors. Gate matches these outputs to the preceding block's hidden-state query. Conv combines current and historical Gate outputs into delta, which is added to the backbone hidden state before attention and FFN computation. Numbered badges identify the six connected stages.}]{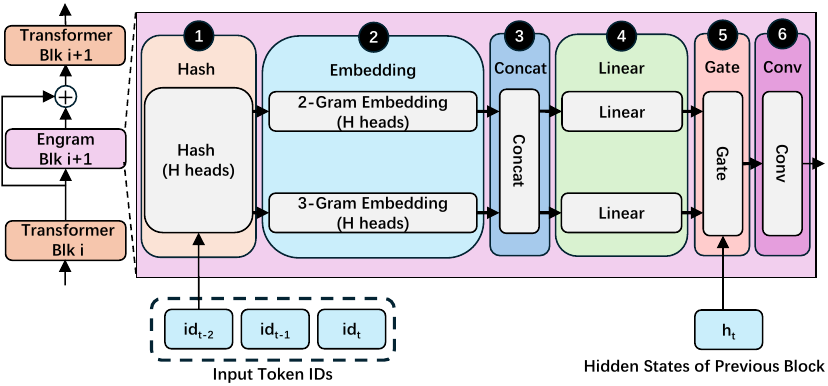}\par}%
\Description{Hash maps token IDs to indices for Embedding; Concat joins the read vectors, and Linear produces key and value vectors. Gate matches these outputs to the preceding block's hidden-state query. Conv combines current and historical Gate outputs into delta, which is added to the backbone hidden state before attention and FFN computation. Numbered badges identify the six connected stages.}
\caption{Engram dataflow from hashing token IDs to the residual update.}
\label{fig:1}
\end{figure}

In summary, this paper makes the following contributions:

\begin{itemize}
\item \textbf{Block-wise disaggregated Engram placement mechanism}. We place embeddings of early Engram blocks in CXL-DRAM and all remaining embeddings in CXL-PCM. HBM retains backbone weights, key/value caches, and working states; host DRAM stages Embedding outputs. Batched Embedding reads, shared access, and reusable pinned buffers support efficient delivery.
\item \textbf{Dependency-aware heterogeneous Engram execution}. We advance hidden-state-dependent Gate--Conv using an earlier hidden state, executing it through CPU-accelerated, GPU-accelerated, or ASIC-accelerated alongside the backbone. Common CUDA fusion reduces residual GPU overhead, and the target adds the completed delta to the backbone hidden state.
\item \textbf{Adaptive minimum lookahead distance scheduler}. We introduce this scheduler, named FlexScheduler, to predict the minimum CPU-accelerated lookahead distance for each workload from Gate--Conv and transfer costs and observed block times.
\end{itemize}

\section{Background}\label{sec:II}

\subsection{LLM Inference Workflow}

Autoregressive inference has prefill and decode phases. Prefill processes input token IDs in parallel and builds per-block key and value caches. Decode reuses and extends these caches to generate one token per sequence per iteration. Prefill is typically compute-bound; decode repeatedly reads weights and cached context for less new computation, making bandwidth and access latency more influential \cite{10609649}.

Within either phase, residual connections add the outputs of attention and FFNs to the backbone hidden state \cite{NIPS2017_3f5ee243}. Successive blocks therefore retain common information while refining the representation \cite{298683}. Consistent with this shared information, ScoutAttention and HybridGen report high similarity between neighboring hidden states \cite{10.1145/3770743.3804320,lin2026hybridgenefficientllmgenerative}.

\subsection{Engram Architecture}

Fig.~\ref{fig:1} shows how six connected Engram stages incorporate stored knowledge into the Transformer backbone. {\Large\ding{182}} \textbf{Hash} forms n-grams from the token IDs of the current token and its predecessors, then applies multiple hash heads to generate indices for each n-gram order. {\Large\ding{183}} \textbf{Embedding} uses these Hash indices to read the corresponding embedding vectors. {\Large\ding{184}} \textbf{Concat} joins the Embedding outputs along the feature dimension to form one memory representation for the token. {\Large\ding{185}} \textbf{Linear} transforms this Concat output into key and value vectors. {\Large\ding{186}} \textbf{Gate} combines these Linear outputs with the preceding block's output hidden state as its query. After normalizing query and key, Gate computes their dot-product compatibility score, converts it into a gating weight, and uses that weight to modulate the corresponding value. {\Large\ding{187}} \textbf{Conv} takes Gate's current output and the same sequence's historical Gate outputs through a short, channel-wise causal convolution. Each sequence maintains its own history. Conv produces delta, which is added to the backbone hidden state before that block's attention and FFN computation. Gate directly depends on the hidden state, and Conv inherits this dependency through Gate's output. Hash, Embedding, Concat, and Linear depend only on the current and preceding token IDs. Once these IDs are known, their results are determined without waiting for intermediate hidden states.

\subsection{CXL Memory Pooling and Sharing}

CXL memory pooling aggregates memory devices through switches, decoupling capacity from individual hosts and allowing hosts to draw capacity on demand instead of provisioning local memory for every peak. CXL memory sharing lets multiple hosts reuse one data copy on devices and topologies that support shared access, reducing replicas \cite{10.1145/3669900,10.1145/3575693.3578835,jain2024memorysharingcxlhardware}.

Such pools can connect memory devices backed by different storage media \cite{10.1145/3582016.3582063}. CXL-DRAM uses DRAM for low-latency access; measurements across four devices report average local latencies of \SIrange{214}{394}{\nano\second} \cite{10.1145/3676641.3715987}. CXL-PCM uses nonvolatile phase-change memory (PCM), offering higher density and potentially lower capacity cost but longer read latency \cite{10.1145/1555754.1555758}. Early device studies estimate two to four times DRAM's density \cite{10.1145/1555754.1555760}.

\section{Motivation}\label{sec:III}

\subsection{Engram Embedding Access Characterization}

Embeddings feature large capacity, fine granularity, and scattered accesses. Public models already have substantial memory capacity: DeepSeek-V4.1-Flash's embeddings occupy about \SI{200}{\giga\byte}, and eight inference nodes storing full replicas require approximately \SI{1600}{\giga\byte}; sharing a static copy reduces this duplicate storage. In our Qwen3-4B configuration, each module has eight 2-gram heads and eight 3-gram heads, with independent embeddings for each head. Each entry stores one 64-element FP32 vector occupying \SI{256}{\byte}; a \SI{4}{\kibi\byte} page holds 16 such entries, and each head's embeddings span about 40,400 pages. Embedding reads one entry per head for each token, giving \SI{4}{\kibi\byte} of retrieved data per Engram despite the large stored capacity. These entries are scattered: an accessed page supplies about \SI{268}{\byte} of useful data on average, or 6.5\% page utilization. Cache-line-granular CXL reads suit the sparse access pattern and can approach local-DRAM latency for retrieval \cite{10.1145/3805621.3807654}.

Placing embeddings involves trade-offs in capacity scaling, access latency, and sharing. HBM offers direct GPU access but limited capacity. Host DRAM expands per-server storage but struggles to keep pace with embeddings growth; retrieved entries still require GPU transfers. CXL pools decouple capacity from individual hosts \cite{280792}. Independent embeddings per Engram block enable tiering: early Engram blocks have short Embedding access windows and use low-latency CXL-DRAM, while later Engram blocks use longer windows to hide CXL-PCM latency and exploit its higher density and potentially lower cost.

\begin{figure}[t!]
{\centering
\includegraphics[width=\linewidth,alt={For each of three sequence lengths, paired stacked bars report CPU and GPU operator latency against batch size. Colors and hatching distinguish Linear, Gate, and Conv. GPU bars are shorter than the paired CPU bars.}]{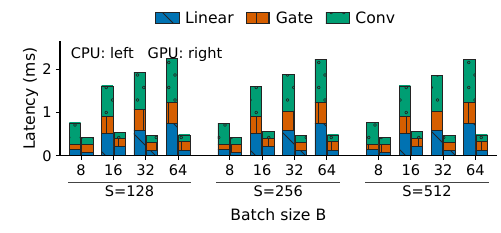}\par}%
\Description{For each of three sequence lengths, paired stacked bars report CPU and GPU operator latency against batch size. Colors and hatching distinguish Linear, Gate, and Conv. GPU bars are shorter than the paired CPU bars.}
\caption{Decode operator times on Qwen3-4B, B=8--64 and S=128,256,512. CPU/GPU form the left/right bars of each pair.}
\label{fig:2}
\end{figure}

\subsection{Engram Computation Characterization}

Linear, Gate, and Conv have different computational characteristics. Linear projects the memory vector into key and value vectors through matrix multiplication. Gate normalizes vectors, computes dot products, and modulates value; Conv performs local multiply-adds independently along each channel \cite{cheng2026conditionalmemoryscalablelookup}.

GPUs suit matrix operations and parallel execution across tokens and channels~\cite{10.1145/1401132.1401152}, enabling faster Linear, Gate, and Conv: for B=16--64, Fig.~\ref{fig:2} shows mean latency reductions of 73.17\%, 56.74\%, and 81.90\% for Linear, Gate, and Conv relative to conventional CPU execution.

Although GPU execution shortens Gate--Conv, running these operations serially at the target block still adds their latency to the backbone's critical path. CPU execution can overlap the GPU backbone, but its window must cover the longer Gate--Conv execution, the hidden-state transfer to the CPU, and the delta transfer back to the GPU. Thus, common CUDA fusion reduces early Engram blocks' GPU overhead. CPU-accelerated speeds up later Gate--Conv, whereas GPU-accelerated runs it on an independent Engram Stream alongside Main Stream.

\subsection{Execution Dependency and Overlap Opportunities}

\begin{figure}[t!]
{\centering
\includegraphics[width=\linewidth,alt={Five curves compare block-input hidden states separated by one to five blocks. Shorter gaps generally have higher cosine similarity through the middle and later blocks. The terminal comparisons include the output state and drop relative to preceding points.}]{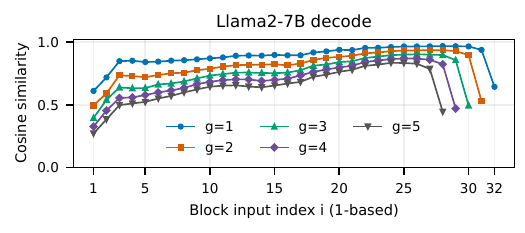}\par}%
\Description{Five curves compare block-input hidden states separated by one to five blocks. Shorter gaps generally have higher cosine similarity through the middle and later blocks. The terminal comparisons include the output state and drop relative to preceding points.}
\caption{Cosine similarity of Llama2-7B decode hidden states \cite{touvron2023llama2openfoundation} at index gaps 1--5. Indices are one-based; terminal points compare against the final output state.}
\label{fig:3}
\end{figure}

Following the six stages in Fig.~\ref{fig:1}, Hash, Embedding, Concat, and Linear depend only on known token IDs and can overlap preceding Transformer computation \cite{cheng2026conditionalmemoryscalablelookup,10.1145/3805621.3807654}. Gate requires the preceding block's output hidden state, and Conv consumes Gate's output. Earlier execution must therefore account for hidden-state differences. Fig.~\ref{fig:3} compares block inputs at different index gaps: a gap of one compares, for example, blocks 10 and 11; a gap of three compares blocks 10 and 13. Middle and later states are generally similar at short gaps, with similarity tending to decrease as the gap grows.

\begin{figure*}
{\centering
\includegraphics[width=.32\linewidth,alt={CPU-accelerated execution. The CPU runs Gate and Conv in fused kernels. FlexScheduler predicts the minimum lookahead distance.}]{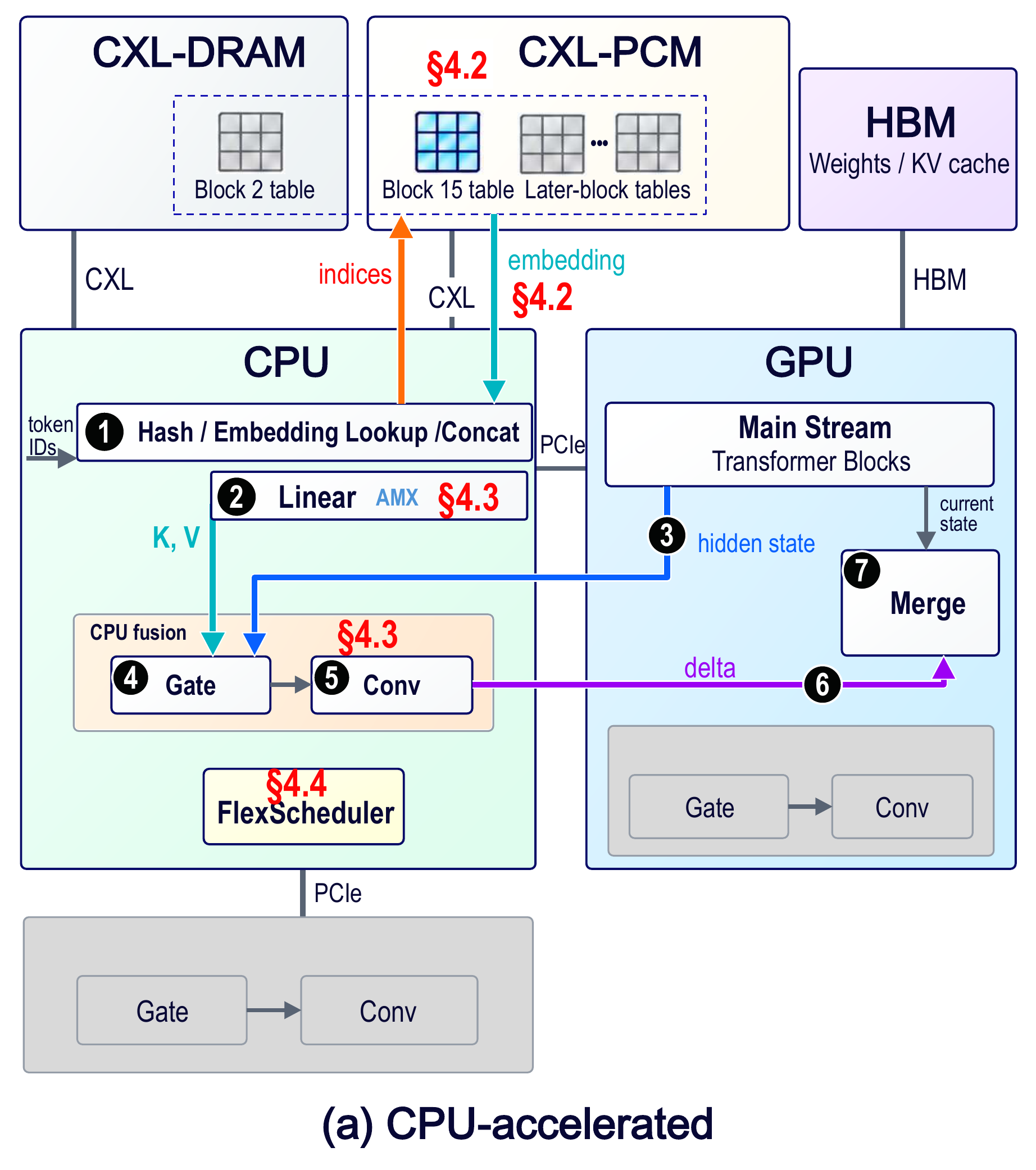}
\hfill
\includegraphics[width=.32\linewidth,alt={GPU-accelerated execution. Gate and Conv run on an independent Engram Stream with one-block lookahead.}]{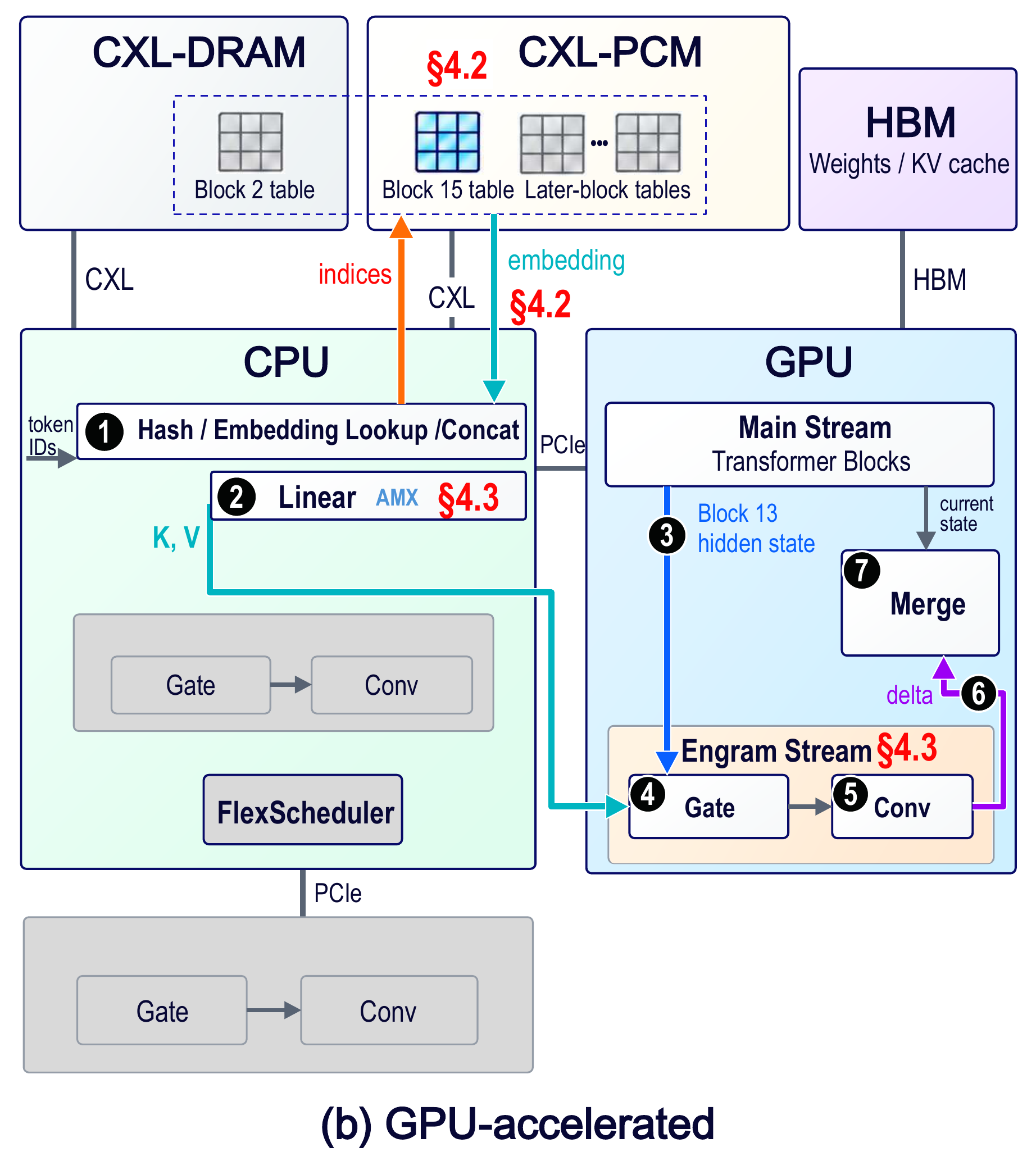}
\hfill
\includegraphics[width=.32\linewidth,alt={ASIC-accelerated execution. A dedicated Engram accelerator runs Gate and Conv with one-block lookahead.}]{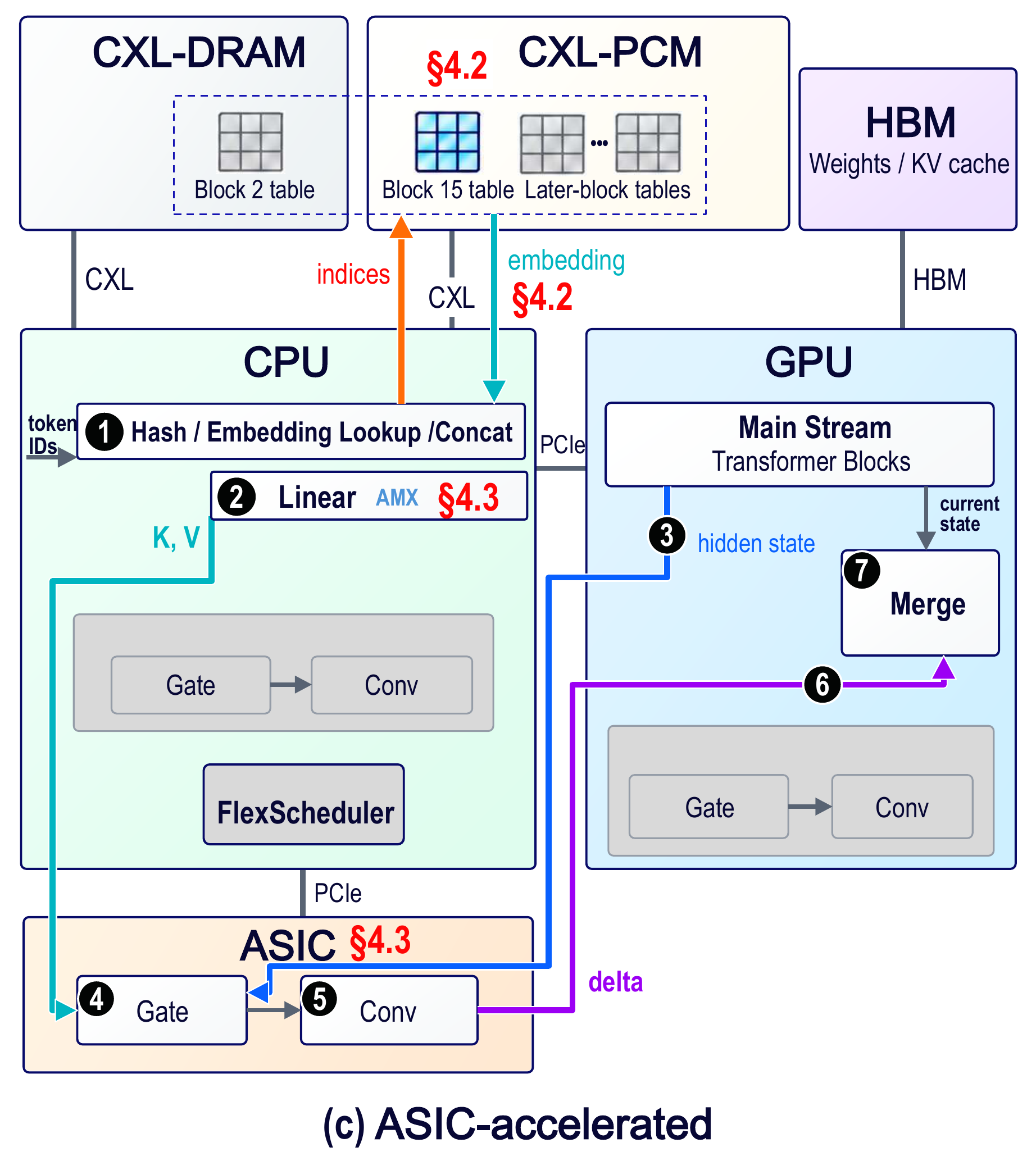}\par}
\caption{Decode execution for Engram block 15: (a) CPU-accelerated, (b) GPU-accelerated, and (c) ASIC-accelerated. FlexScheduler predicts the minimum lookahead distance for (a); (b) and (c) use $d=1$. Numbered black badges mark dataflow, red labels link design subsections, and gray components are inactive.}
\label{fig:4}
\Description{Each panel follows token IDs through Hash, Embedding, Concat, Linear, hidden-state input, Gate, Conv, and addition of delta to the backbone hidden state. The CPU-accelerated panel uses fused kernels, the GPU-accelerated panel an independent Engram Stream, and the ASIC-accelerated panel a dedicated Engram accelerator. Gray components are inactive; FlexScheduler predicts the minimum CPU-accelerated lookahead distance; ASIC-accelerated uses fixed one-block lookahead.}
\end{figure*}

These curves motivate an accuracy--latency trade-off: increasing lookahead provides more backbone time for Gate and Conv but increases hidden-state differences from the target block input. Engram block 2 has only one preceding block, limiting overlap of Embedding, Gate, and Conv. Engram block 15 has a longer preceding sequence and several candidate hidden states. CPU-accelerated must therefore account for Hash, Embedding, Concat, and Linear readiness, then compare Gate--Conv and transfer costs with the intervening Transformer blocks' execution time to predict the minimum lookahead distance. Any unfinished Gate--Conv execution or transfer at the target becomes backbone waiting time.

\section{FlexEngram Design}\label{sec:IV}

\subsection{System Overview}

FlexEngram's tiered hierarchy keeps backbone state in HBM, stages Embedding outputs in host DRAM, and places early Engram blocks' embeddings in CXL-DRAM and the remaining embeddings in CXL-PCM. Alongside this placement, executing Gate--Conv before the target block reduces backbone waiting. The placement and access mechanisms (\textbf{\S{}4.2}: \hyperref[par:placement]{placement} and \hyperref[par:retrieval]{access}) support heterogeneous execution (\textbf{\S{}4.3}), while FlexScheduler predicts the minimum CPU-accelerated lookahead distance (\textbf{\S{}4.4}).

\begin{figure}[t!]
{\centering
\includegraphics[width=0.90\linewidth,alt={The hierarchy separates HBM backbone state, DRAM staging, early CXL-DRAM embeddings, and later CXL-PCM embeddings. The entry enlargement shows sixteen 256 B entries per page, each with 64 FP32 elements. Yellow highlights identify entries selected by one token.}]{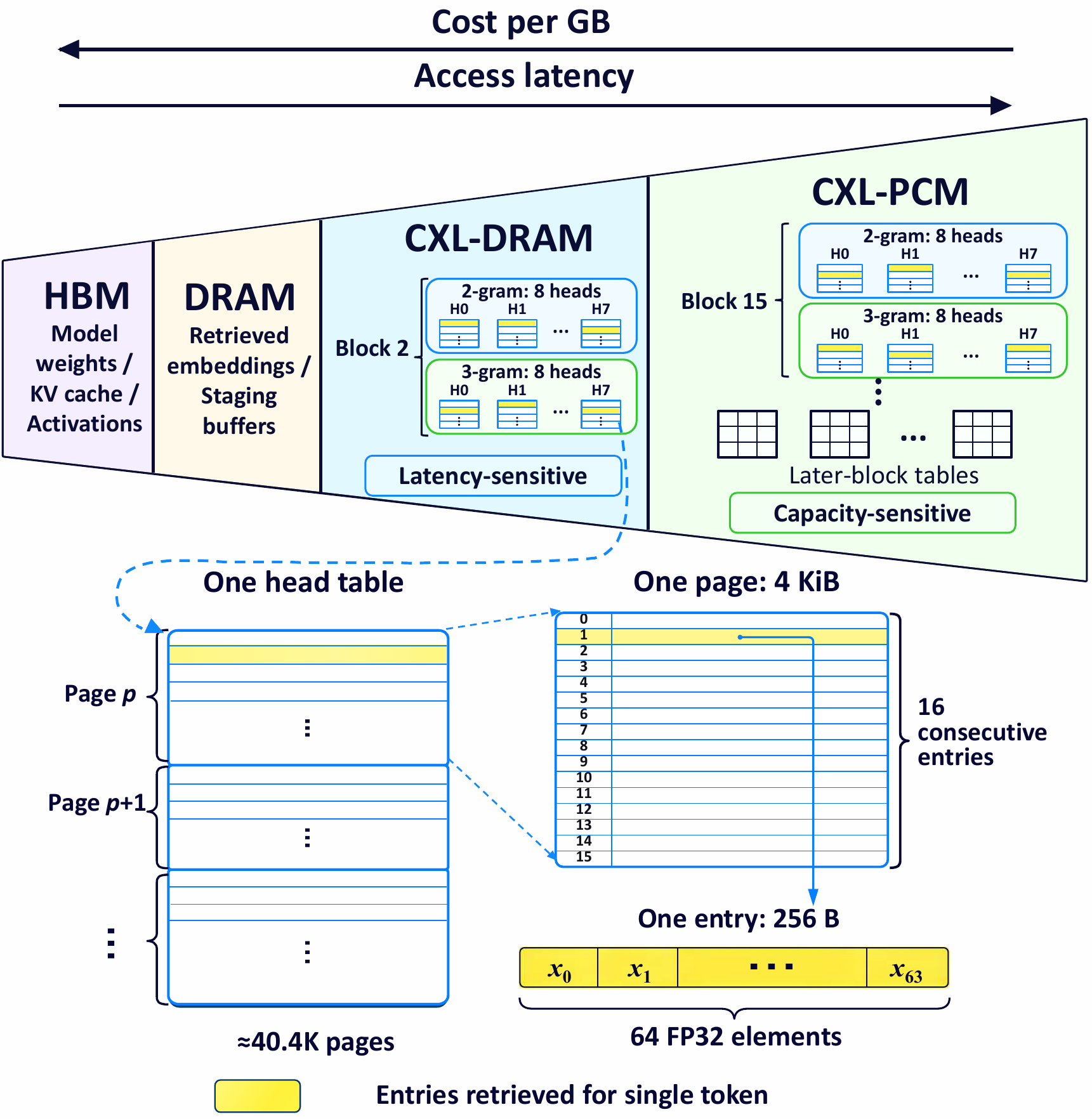}\par}%
\Description{The hierarchy separates HBM backbone state, DRAM staging, early CXL-DRAM embeddings, and later CXL-PCM embeddings. The entry enlargement shows sixteen 256 B entries per page, each with 64 FP32 elements. Yellow highlights identify entries selected by one token.}
\caption{Placement of independent block embeddings and their entry layout. Yellow entries belong to one token Embedding access.}
\label{fig:5}
\end{figure}

Fig.~\ref{fig:4}(a)--(c) follows Engram block 15 after Engram block 2 completes. The CPU executes Hash, Embedding, and Concat ({\Large\ding{182}}), then AMX-accelerated Linear produces key and value ({\Large\ding{183}}, \textbf{\S{}4.3}, \hyperref[par:cpu]{CPU acceleration}). Once an earlier hidden state ({\Large\ding{184}}) is available, the chosen execution path computes Gate ({\Large\ding{185}}) and Conv ({\Large\ding{186}}). It delivers delta ({\Large\ding{187}}) for addition to the backbone hidden state at block 15 ({\Large\ding{188}}). FlexScheduler predicts the minimum CPU-accelerated lookahead distance; GPU-accelerated and ASIC-accelerated use block 13 output to overlap block 14. All three final paths apply CUDA operator optimizations to Gate and Conv of Engram block 2 on the GPU (\textbf{\S{}4.3}, \hyperref[par:cuda]{CUDA fusion}).

\subsection{Tiered Engram Storage}

\runinheading{Embedding placement and data layout}\phantomsection\label{par:placement} FlexEngram places embeddings of early Engram blocks in CXL-DRAM and those of all remaining Engram blocks in CXL-PCM (Fig.~\ref{fig:5}). HBM holds backbone weights, key and value caches, and activations, while host DRAM buffers Embedding outputs. Engram block 2 has little preceding computation and low early-state similarity, so its embeddings use CXL-DRAM. Embeddings of all remaining Engram blocks use denser, lower-cost CXL-PCM, hiding its longer access latency behind preceding block computation.

\runinheading{Efficient Embedding access and transfer}\phantomsection\label{par:retrieval} To deliver entries from either CXL medium, Adaptive Retrieval uses a common path that writes selected entries directly into reusable contiguous pinned buffers, avoiding repeated allocation and copying. Ready, compatible requests for one GPU are read as a batch and delivered separately; batching never waits for additional arrivals. After each transfer completes, the chosen execution path can reuse the buffer. When readers contend for shared CXL bandwidth \cite{10.1109/DAC63849.2025.11132607}, coordination is enabled only when the predicted permit wait plus Embedding read time is shorter than direct reading. The coordination mechanism limits concurrency and prioritizes clients that have received less service. Each client then reads its selected entries and transfers them to its local GPU.

\subsection{Dependency-Aware Heterogeneous Engram Execution}

\runinheading{Dependency-aware early execution} Hash, Embedding, Concat, and Linear produce key and value from token IDs before consumption. Gate compares the earlier block's hidden state with key and uses their match to weight value; Conv uses Gate's output. Using this earlier hidden state lets Gate--Conv overlap subsequent blocks. The target adds only delta to the backbone hidden state, preserving the backbone's in-order computation.

\begin{figure}
{\centering
\includegraphics[width=\linewidth,alt={Six timelines show memory, CPU, and GPU work for two baselines, prefill, and three decode execution paths. Hash, Embedding, Concat, Linear, Gate, and Conv overlap preceding or subsequent Transformer computation in the accelerated paths. The GPU-accelerated timeline separates Main Stream and Engram Stream; red completion references permit comparison with PoolingEngram.}]{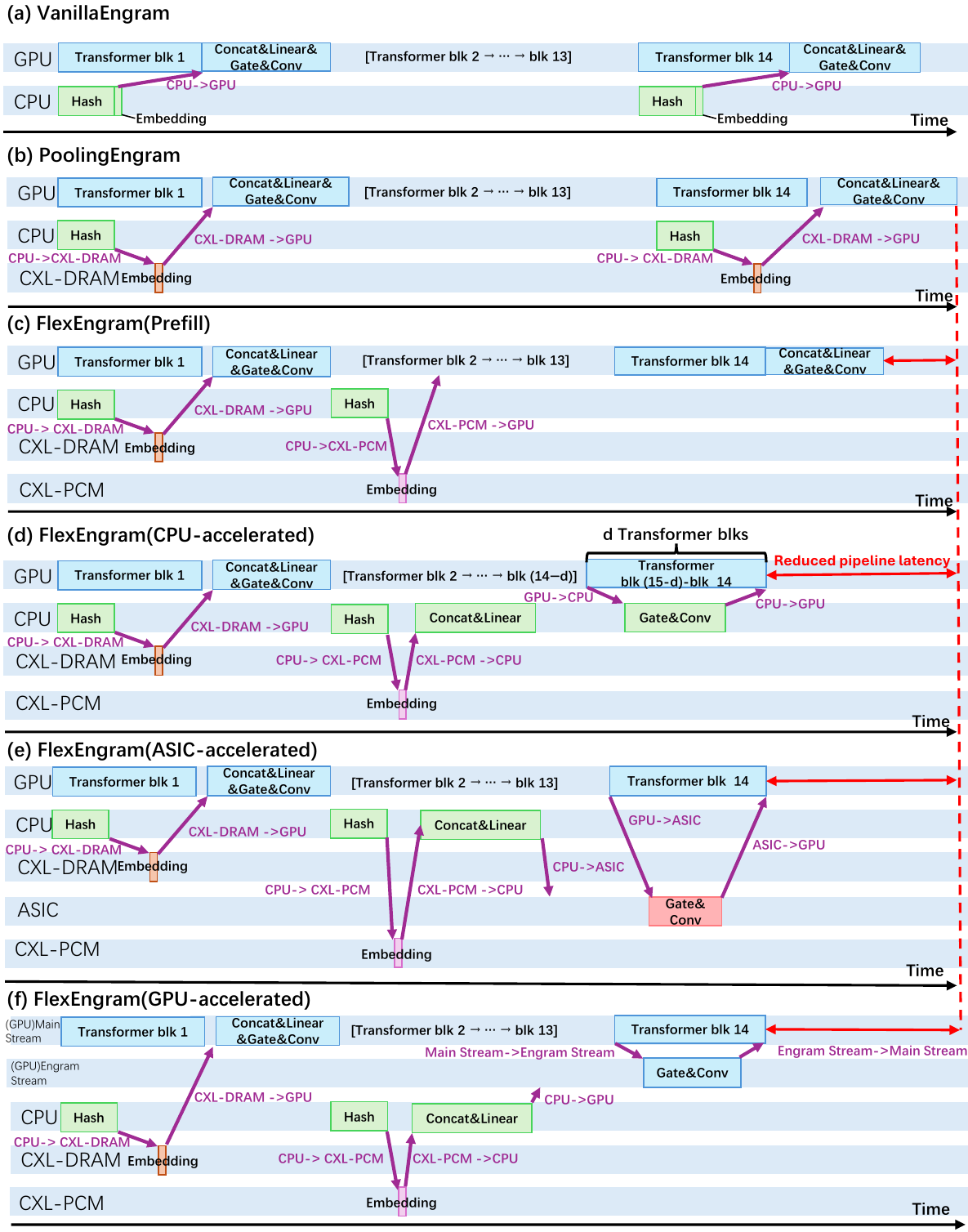}\par}%
\Description{Six timelines show memory, CPU, and GPU work for two baselines, prefill, and three decode execution paths. Hash, Embedding, Concat, Linear, Gate, and Conv overlap preceding or subsequent Transformer computation in the accelerated paths. The GPU-accelerated timeline separates Main Stream and Engram Stream; red completion references permit comparison with PoolingEngram.}
\caption{Execution timelines for (a) VanillaEngram, (b) PoolingEngram, (c) prefill, and (d)--(f) CPU-accelerated, ASIC-accelerated, and GPU-accelerated decode. Only (d) uses FlexScheduler; (e) and (f) use one-block lookahead. All three fuse early Engram GPU operations. Red dashed lines mark PoolingEngram completion; arrows show reductions. Durations are schematic.}
\label{fig:6}
\end{figure}

Fig.~\ref{fig:6}(a)--(c) show VanillaEngram~\cite{cheng2026conditionalmemoryscalablelookup}, PoolingEngram~\cite{10.1145/3805621.3807654}, and prefill. Their embeddings reside in host DRAM, shared CXL, and CXL-DRAM/CXL-PCM, respectively. The CPU executes Hash and initiates Embedding access, while Concat, Linear, Gate, and Conv execute on the GPU at the target block. Fig.~\ref{fig:6}(d)--(f) show CPU-accelerated, ASIC-accelerated, and GPU-accelerated decode, respectively. After Engram block 2 completes, Hash, Embedding, Concat, and Linear execute early for Engram block 15 to produce key and value. Gate and Conv then use an earlier block's hidden state and run in fused CPU kernels, on the dedicated Engram accelerator, or on the Engram Stream while subsequent blocks run on the backbone. FlexScheduler predicts the minimum CPU-accelerated lookahead distance, whereas ASIC-accelerated and GPU-accelerated use fixed one-block lookahead. CPU-accelerated and ASIC-accelerated return delta to the GPU, and GPU-accelerated joins the two streams. At the target block, all three add delta to the backbone hidden state. Red arrows compare completion with PoolingEngram.

\runinheading{CUDA kernel fusion}\phantomsection\label{par:cuda} For Engram operations retained on the GPU, we fuse two groups of operations: Gate postprocessing and Conv. The first combines the nonlinear transformation of Gate scores with value modulation, applying the weights directly to the corresponding token channels. The second combines channel-wise causal Conv, SiLU \cite{ELFWING20183}, residual addition, casting, and scaling to reduce intermediate-tensor writebacks, rereads, and kernel launches. Linear projection, normalization, and dot products retain the original operators; the fused parts preserve the original computation and conversion order.

\runinheading{CPU-based Engram acceleration}\phantomsection\label{par:cpu} We first concatenate the Linear weight matrices for key and value along the output dimension into one A\(\times\)2H matrix, where A and H are the Concat output and hidden dimensions. Using oneDNN with AMX, we multiply the Concat output by this combined weight matrix once, then split the result along the output dimension to obtain key and value. This avoids separate Linear calls and reduces repeated input reads.

We then execute Gate and Conv with one CPU thread team. This reduces thread switches and startup costs. Small batches use fewer threads, whereas large batches expose more parallelism. Linear reuses BF16 weights, with BF16 Gate--Conv I/O and FP32 reduction and Conv intermediates. Reusable working and pinned output buffers return delta to the GPU, signaled by a readiness event.

\begin{figure}
{\centering
\includegraphics[width=\linewidth,alt={A sequence router distributes complete rows from the 1024-bit input to four engines with 32 lanes each and private history. Output arbitration joins their results and completion signals. The engine detail connects normalization and dot reduction to signed square root, sigmoid, value modulation, and a four-tap history convolution.}]{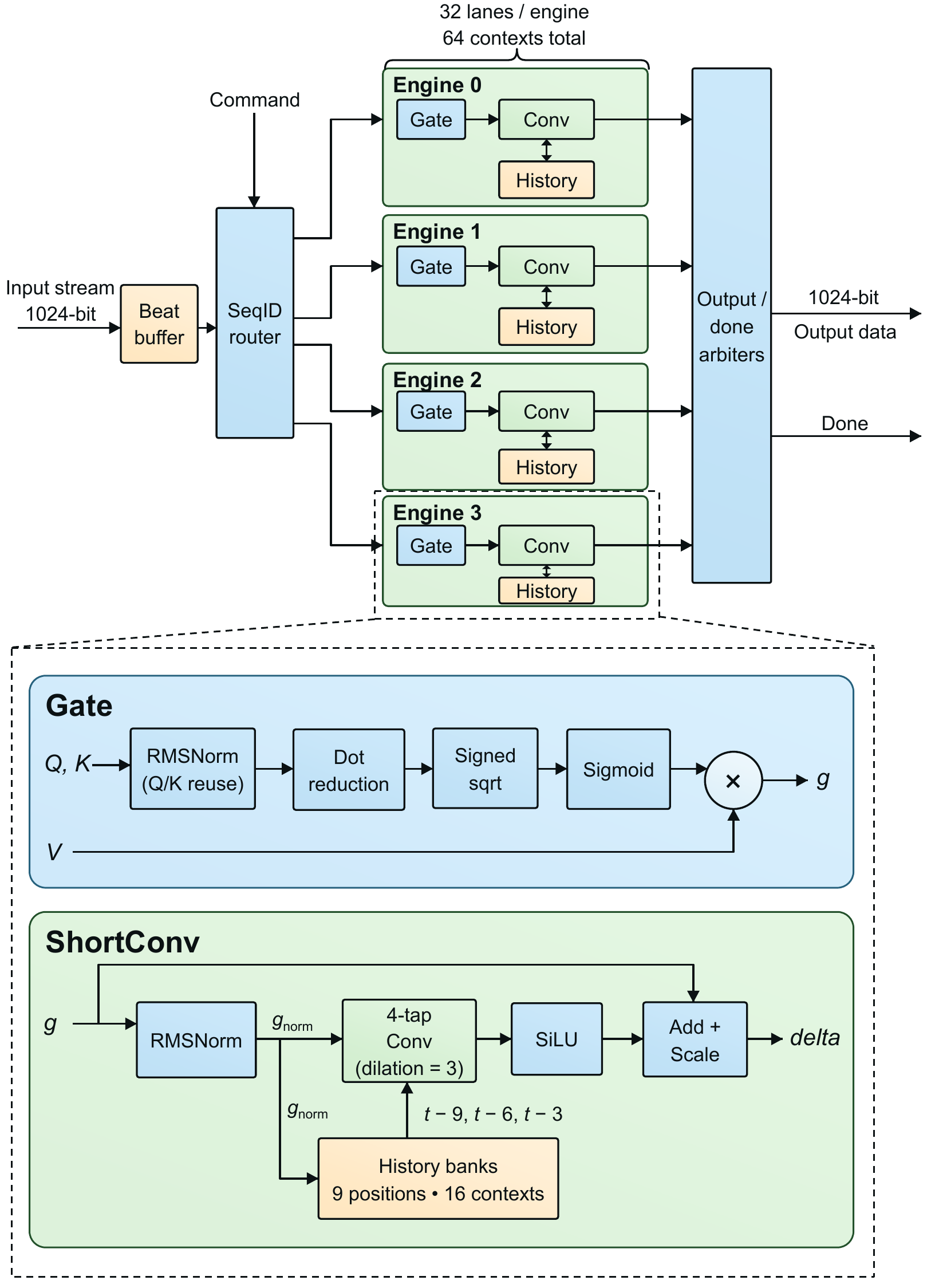}\par}%
\Description{A sequence router distributes complete rows from the 1024-bit input to four engines with 32 lanes each and private history. Output arbitration joins their results and completion signals. The engine detail connects normalization and dot reduction to signed square root, sigmoid, value modulation, and a four-tap history convolution.}
\caption{Dedicated Engram accelerator, with the four-engine unit above and one engine datapath below.}
\label{fig:7}
\end{figure}

\runinheading{Dedicated Engram accelerator architecture} A unit routes complete token rows by sequence ID to four engines (Fig.~\ref{fig:7}). Each has 32 hidden-dimension lanes and supports dimensions up to 4096. The CPU executes Linear for key and value and handles codecs/control; the dedicated Engram accelerator executes Gate--Conv with sequence and channel parallelism.

Gate shares root-mean-square normalization (RMSNorm) hardware between the hidden state and key \cite{NEURIPS2019_1e8a1942}, using registered trees for square sums and dot products. Scaling, signed square root, and LUT sigmoid weight the value. ShortConv combines the normalized current value with history offsets 3, 6, and 9: four taps at dilation=3, followed by SiLU, residual addition, and scaling.

History is banked by channel and time, retaining nine positions. Each engine holds 16 contexts, giving 64 per unit under a 6-bit sequence ID (SID). A \SI{1024}{\bit} stream supplies hidden state, key, value, and token-specific normalization/convolution parameters to row buffers. Output arbitration transmits each token contiguously before signaling completion. A later Engram uses one/two/four units at $B\leq64$/$B=128$/$B=256$, sharing host I/O teams and dividing bidirectional link bandwidth among active units.

Because each Engram block needs independent Conv history, the pair (Engram block, sequence) identifies a resident context, with consecutive groups of 64 mapped to units. At B=64 with eight Engrams, Engram block 2 remains on the GPU and the seven later Engram blocks occupy 448 contexts across seven units. Each replicated unit includes its compute logic and history SRAM.

\runinheading{GPU-accelerated execution} GPU-accelerated places Gate--Conv on an added Engram Stream alongside the backbone's Main Stream (Fig.~\ref{fig:6}(f)). Engram Stream replays Engram block 15 Gate--Conv using block 13 output while Main Stream computes block 14; a stream event then signals that delta is ready to be added to the backbone hidden state at block 15. Keeping inputs and results on the GPU eliminates host round trips.

\subsection{FlexScheduler}

\begin{figure}
{\centering
\includegraphics[width=\linewidth,alt={Configuration and calibration feed the CPU workload and CPU-stage latency model. The critical-path estimate and the mean time of completed blocks feed the bounded minimum-distance calculation, which outputs the lookahead distance and launch block.}]{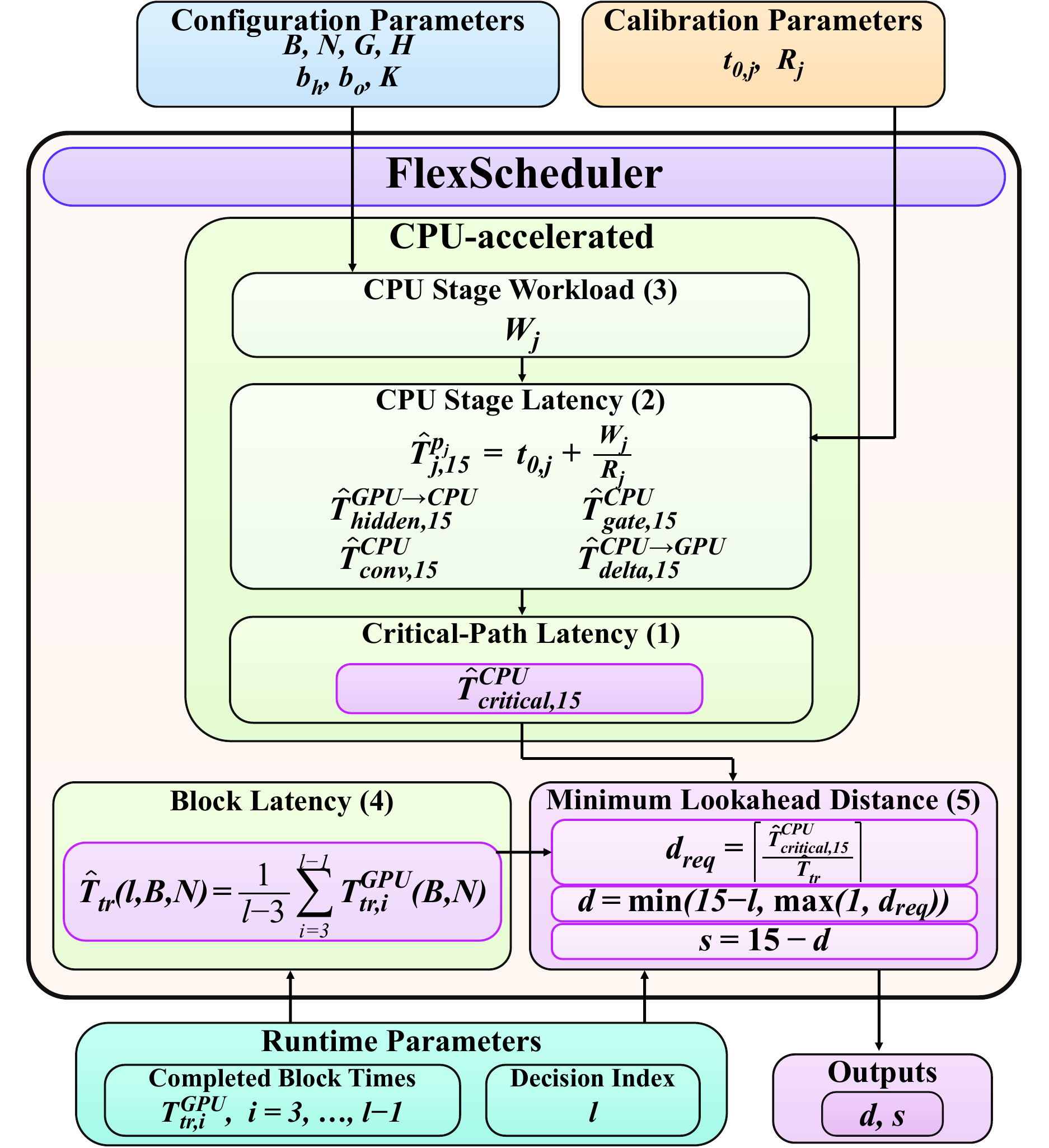}\par}%
\Description{Configuration and calibration feed the CPU workload and CPU-stage latency model. The critical-path estimate and the mean time of completed blocks feed the bounded minimum-distance calculation, which outputs the lookahead distance and launch block.}
\caption{FlexScheduler for CPU-accelerated. Configuration, calibration, and completed-block timings support prediction of the minimum lookahead distance $d$ and computation of the corresponding block $s$. Parenthesized numbers refer to equations.}
\label{fig:8}
\end{figure}

FlexScheduler predicts the minimum CPU-accelerated lookahead distance from Gate--Conv and transfer costs and completed-block timings (Fig.~\ref{fig:8}). In CPU-accelerated, AMX accelerates Linear driven by token IDs, while Gate and Conv execute in fused CPU kernels. The scheduler predicts this path's total latency and compares it with the GPU compute time available within the lookahead window. Workload configuration determines the amount of work, and calibration supplies startup costs and effective service rates. The scheduler uses completed-block timings to predict the overlap time available from each subsequent block. 

\runinheading{Critical-path latency} The budget $\widehat T_{\mathrm{critical},15}^{\mathrm{CPU}}$ sums predicted GPU-to-CPU transfer of the earlier hidden state, Gate, Conv, and delta CPU-to-GPU return, in order (Equation~\eqref{eq:1}). Hash, Embedding, Concat, and Linear depend only on token IDs and must supply key/value before Gate--Conv; this sets input readiness, adding no budget terms. The critical-path CPU superscript denotes CPU-accelerated; stage superscripts denote location or transfer direction. Times use \si{\milli\second}:

\begin{equation*}\begin{aligned}
\widehat T_{\mathrm{critical},15}^{\mathrm{CPU}}
 &=\widehat T_{\mathrm{hidden},15}^{\mathrm{GPU}\rightarrow\mathrm{CPU}}
   +\widehat T_{\mathrm{gate},15}^{\mathrm{CPU}}\\
 &\quad+\widehat T_{\mathrm{conv},15}^{\mathrm{CPU}}
   +\widehat T_{\mathrm{delta},15}^{\mathrm{CPU}\rightarrow\mathrm{GPU}}.
\end{aligned}
\tag{1}\label{eq:1}\end{equation*}

\runinheading{CPU stage latency} Equation~\eqref{eq:2} models CPU stages and host--GPU transfers as fixed startup plus CPU stage workload divided by a calibrated rate. For stage $j$, $p_j$ denotes location or direction; $t_{0,j}$ startup, $W_j$ workload, and $R_j$ effective rate (\si{\byte\per\milli\second}). Stage-specific calibration captures actual CPU-kernel and transfer-path startup and work-dependent costs:

\begin{equation*}\begin{gathered}
\widehat T_{j,15}^{p_j}=t_{0,j}+\frac{W_j}{R_j},\\
(p_j,j)=(\mathrm{GPU}\!\rightarrow\!\mathrm{CPU},\mathrm{hidden}),\ 
(\mathrm{CPU},\mathrm{gate}),\\
(\mathrm{CPU},\mathrm{conv}),\ 
(\mathrm{CPU}\!\rightarrow\!\mathrm{GPU},\mathrm{delta}).
\end{gathered}
\tag{2}\label{eq:2}\end{equation*}

\runinheading{CPU stage workload} Equation~\eqref{eq:3} counts decode-kernel logical byte work. $B,N,G,H$ denote batch size, tokens per sequence, branches, and hidden dimension; $b_h,b_o$ denote input/output bytes per element. Decode uses one token per sequence. With $N=G=1$ and $b_h=b_o=2$ bytes, hidden input and delta return each carry $2BH$ bytes. For four-tap Conv ($K=4$), elementwise Gate and Conv work is $16BH$ and $28BH$, respectively. Gate parameter reads and BF16-to-FP32 conversion add $6H(K+2)$; writing Conv's FP32 token-normalization coefficients adds $4B$:

\begin{equation*}\begin{aligned}
W_{\mathrm{hidden}}&=BNGHb_h,\qquad \mathord{W_{\mathrm{delta}}=BNGHb_o},\\
W_{\mathrm{gate}}&=16BH+6H(K+2),\\
W_{\mathrm{conv}}&=28BH+4B.
\end{aligned}
\tag{3}\label{eq:3}\end{equation*}

\runinheading{Block latency} We predict the overlap time available from each subsequent GPU block using completed-block timings. Let $l$ denote the decision position, with $l\geq4$; $T_{\mathrm{tr},i}^{\mathrm{GPU}}$ is block $i$'s measured time. Equation~\eqref{eq:4} averages blocks 3 through $l-1$ (a total of $l-3$ samples), excluding the atypical first two:

\begin{equation*}\widehat T_{\mathrm{tr}}(l,B,N)
=\frac{1}{l-3}\sum_{i=3}^{l-1}T_{\mathrm{tr},i}^{\mathrm{GPU}}(B,N).
\tag{4}\label{eq:4}\end{equation*}

\runinheading{Minimum lookahead distance} The CPU-accelerated critical path divided by mean block time, rounded upward, gives required overlap in blocks, $d_{\mathrm{req}}$. Equation~\eqref{eq:5} gives the range of $d$ as $[1,15-l]$, with corresponding block index $s=15-d$. When the required window fits the available window, this is the smallest distance predicted to cover execution:

\begin{equation*}\begin{aligned}
d_{\mathrm{req}}
 &=\left\lceil\frac{\widehat T_{\mathrm{critical},15}^{\mathrm{CPU}}}
                         {\widehat T_{\mathrm{tr}}}\right\rceil,\\
d&=\min\left(15-l,\mathord{\max(1,d_{\mathrm{req}})}\right),\\
s&=15-d.
\end{aligned}
\tag{5}\label{eq:5}\end{equation*}

The chosen launch uses the output of block $s-1$ as Gate's hidden-state input while the GPU continues blocks $s$ through 14. At $d=3$, block 11 output starts Gate--Conv at block 12, providing the block 12--14 window; at $d=1$, block 13 output serves as Gate's hidden-state input during block 14. The key and value must also be ready before Gate and Conv begin. At block 15, the target adds delta to the backbone hidden state.

\begin{figure*}
{\centering
\includegraphics[width=\linewidth,alt={Three Qwen panels plot candidate agreement against lookahead and normalized throughput. Recall@32 shows an overall decrease with lookahead. CPU-accelerated throughput reaches 0.979 at distance two from 0.933 at zero and plateaus within 0.979 to 0.989. The conditional ASIC-accelerated series equals one for distances one through seven, with an estimated zero-distance point.}]{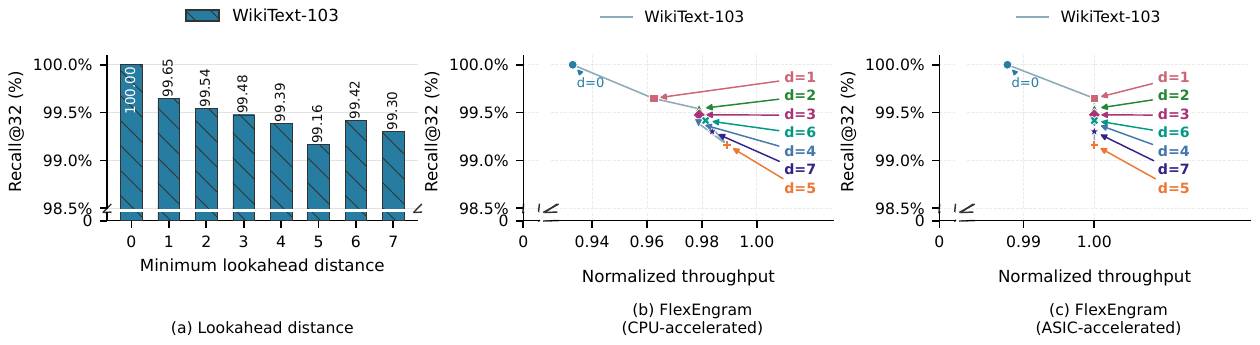}\par}%
\Description{Three Qwen panels plot candidate agreement against lookahead and normalized throughput. Recall@32 shows an overall decrease with lookahead. CPU-accelerated throughput reaches 0.979 at distance two from 0.933 at zero and plateaus within 0.979 to 0.989. The conditional ASIC-accelerated series equals one for distances one through seven, with an estimated zero-distance point.}
\caption{Lookahead on Qwen3-4B, WikiText-103, B=32, S=256: (a) Recall@32, (b) measured CPU-accelerated throughput, (c) ASIC-accelerated throughput. Throughput uses block-2-only Engram as the reference.}
\label{fig:9}
\end{figure*}

\section{Evaluation}\label{sec:V}

\subsection{Experimental Setup}

\runinheading{Platforms and workloads} To measure decode performance, we evaluate Qwen3-4B \cite{yang2025qwen3technicalreport} and Llama-3.1-8B-Instruct on WikiText-103 \cite{merity2017pointer}: fifteen valid loads each, B=8--256, S=128--512 (Table~\ref{tab:1}).

\runinheading{Baselines and comparison schemes} Two-module runs use Engram blocks 2/15. Complete FlexEngram uses CXL-DRAM/CXL-PCM and CUDA fusion, with FlexScheduler for CPU-accelerated and fixed one-block lookahead for GPU-accelerated and ASIC-accelerated.
\begin{itemize}
\item \textbf{VanillaEngram} \cite{cheng2026conditionalmemoryscalablelookup} / \textbf{PoolingEngram} \cite{10.1145/3805621.3807654}: GPU execution with host-DRAM / GPU-direct CXL embeddings.
\item \textbf{KernelEngram}: Kernel-Smith weight precomputation and fused Triton operators in VanillaEngram \cite{du2026kernelsmithunifiedrecipeevolutionary,10.1145/3315508.3329973}.
\item \textbf{CXL + GPU} / \textbf{CXL + fixed CPU}: GPU execution / conventional CPU Gate--Conv one block early.
\item \textbf{FlexEngram}: Section~\ref{sec:IV}'s three execution paths. \textbf{No Engram} removes modules and embeddings.
\end{itemize}

\begin{table}[htbp]
\centering
\caption{Hardware and memory configuration.}
\label{tab:1}
\begin{tabular}{p{.21\linewidth}p{.66\linewidth}}
\toprule
\textbf{Item} & \textbf{Configuration} \\
\midrule
CPU & 2 \(\times\) Xeon Platinum 8468V (SPR); 48 cores/socket, \SI{2.9}{\giga\hertz}; SMT, AMX-BF16 \\
Caches & Per socket: \SI{96}{\mebi\byte} L2, \SI{97.5}{\mebi\byte} LLC \\
OS & Linux v6.6.0vtism-patch+ \\
GPU & 1 \(\times\) NVIDIA A100-SXM4, \SI{80}{\giga\byte} HBM2e; \SI{2039}{\giga\byte\per\second}; PCIe 4.0 \(\times\)16 \\
ASIC link & \SI{28}{\giga\byte\per\second}/direction, aggregate shared \\
\midrule
Host DRAM & DDR5; \SI{220}{\gibi\byte} total (\SI{189}{\gibi\byte} on node 1); NUMA 0, 1; integrated controller \\
CXL-DRAM & CXL 2.x; 63.0 + \SI{59.0}{\gibi\byte}; NUMA 3, 4; Montage M88MX5891 \\
CXL-PCM & CXL 2.x / PCIe 5.0 \(\times\)8; \SI{42.6}{\gibi\byte}; NUMA 2; PCI ID 2042:0ddb \\
\bottomrule
\end{tabular}
\end{table}

\runinheading{Metrics} Throughput measures generated tokens/s; relative gains equally weight matched loads, models, and paths. For each target Engram block, we measure consumption latency as the GPU idle time between the end of Transformer computation in the preceding block and the start of Transformer computation in the target block.

Because the required pretrained Engram weights are unavailable, we use randomly initialized Engram to evaluate lookahead consistency. We report the following metrics:

\begin{itemize}
\item \textbf{Recall@32.} With identical token IDs and parameters, we compare the LLM's end-to-end outputs under original ($d=0$) and advance-$d$ execution. At each position, we take the intersection of the two top-32 candidate-token sets, divide its size by 32, and average across all positions. Identical sets at every position give 100\%.

\item \textbf{GSM8K accuracy.} We use GSM8K~\cite{cobbe2021trainingverifierssolvemath} to evaluate the accuracy loss from early execution separately from the accuracy gains of incorporating Engram. Early execution affects only Gate's query. Following Engram's weighted update \cite{cheng2026conditionalmemoryscalablelookup}, our residual proxy changes the pretrained block output from $h+\Delta$ to $h+\alpha_d\Delta$, where $\Delta$ is the value added by the Transformer block to the backbone hidden state. We compute $\alpha_d=\operatorname{clip}(\cos(q_d,h),0,1)$. Here $h$ is the target input hidden state and $q_d$ is the hidden state from $d$ blocks earlier. At $d=0$, $\alpha_0=1$ restores the original output. Each configuration maintains independent hidden-state and key/value cache trajectories, leaves unselected blocks/prefill unchanged, and runs the full test set once: early $d=0$, later $d=1,2,3$.
\end{itemize}

\subsection{FlexScheduler Validation}

\begin{table}
\centering
\caption{CPU-accelerated times on Llama. Xfer: hidden input + delta return; G+C: Gate + Conv; Total: critical path. block: predicted blocks 3--8 mean / measured window mean; $d_p/d_r$: predicted/actual distance.}
\label{tab:2}
\begin{tabular}{lccccc}
\toprule
& \multicolumn{4}{c}{\textbf{predict/real (ms)}} & \\
\raisebox{0.5\normalbaselineskip}[0pt][0pt]{\textbf{B$\times$S}} & \textbf{Xfer} & \textbf{G+C} & \textbf{Total} & \textbf{block} & \raisebox{0.5\normalbaselineskip}[0pt][0pt]{\textbf{$d_p/d_r$}} \\
\midrule
8$\times$128 & .06/.07 & .43/.34 & .49/.41 & .58/.57 & 1/1 \\
8$\times$256 & .06/.05 & .43/.37 & .49/.42 & .59/.58 & 1/1 \\
8$\times$512 & .06/.07 & .43/.38 & .49/.45 & .61/.62 & 1/1 \\
16$\times$128 & .07/.06 & .82/.71 & .89/.77 & .65/.67 & 2/2 \\
16$\times$256 & .07/.07 & .82/.78 & .89/.85 & .68/.70 & 2/2 \\
16$\times$512 & .07/.07 & .82/.72 & .89/.79 & .75/.77 & 2/2 \\
32$\times$128 & .08/.08 & 1.60/1.68 & 1.68/1.76 & .77/.80 & 3/3 \\
32$\times$256 & .08/.08 & 1.60/1.70 & 1.68/1.78 & .84/.87 & 3/3 \\
32$\times$512 & .08/.08 & 1.60/1.59 & 1.68/1.67 & .98/1.00 & 2/2 \\
64$\times$128 & .10/.11 & .96/.98 & 1.06/1.08 & 1.02/1.04 & 2/2 \\
64$\times$256 & .10/.10 & .96/.97 & 1.06/1.08 & 1.16/1.16 & 1/1 \\
64$\times$512 & .10/.10 & .96/.94 & 1.06/1.04 & 1.44/1.43 & 1/1 \\
128$\times$128 & .15/.15 & 1.83/1.77 & 1.98/1.92 & 1.65/1.65 & 2/2 \\
128$\times$256 & .15/.15 & 1.83/1.81 & 1.98/1.96 & 1.93/1.93 & 2/2 \\
256$\times$128 & .24/.24 & 3.57/3.58 & 3.81/3.82 & 2.91/2.90 & 2/2 \\
\bottomrule
\end{tabular}
\end{table}

\runinheading{Workload-adaptive lookahead} We first compare the predicted minimum lookahead distance with the minimum distance measured for each workload. At B=32, increasing S from 128 to 512 leaves the predicted critical path near \SI{1.685}{\milli\second} but raises predicted block time from 0.772 to \SI{0.976}{\milli\second}. Each block therefore provides more time to cover the same critical path, reducing the minimum lookahead from three blocks to two. To track this workload-dependent overlap requirement, FlexScheduler combines predicted Gate--Conv and transfer costs for CPU-accelerated with completed GPU-block timings to predict the minimum lookahead distance. Due to space limitations, Table~\ref{tab:2} shows validation results for Llama-3.1-8B-Instruct only; predicted and measured minimum distances agree across all fifteen loads (100\%).

\subsection{Accuracy--Performance Trade-off}

\runinheading{Recall@32} We next vary $d$ to quantify the trade-off between accuracy and lookahead distance. Fig.~\ref{fig:9}(a) shows an overall decrease in output consistency as $d$ increases, indicating greater lookahead-induced accuracy loss in this Recall@32 comparison.

\runinheading{Normalized throughput and Recall@32} We compare saturation points. CPU-accelerated plateaus at $d=2$ (0.979--0.989); the ASIC-accelerated estimate plateaus at $d=1$ (1.000; Fig.~\ref{fig:9}(b), (c)). Their Recall@32 scores are 99.54\% and 99.65\%, respectively. The latter needs less overlap and retains greater consistency with its shorter lookahead.

\subsection{Ablation Study}

\begin{figure*}
{\centering
\includegraphics[width=\linewidth,alt={Six grouped-bar panels compare ablations by batch size, using rows for Qwen and Llama and columns for sequence length. The dashed VanillaEngram reference is one. Colors and hatching identify execution paths and ablations; marked axis breaks preserve zero and OOM labels identify unavailable loads.}]{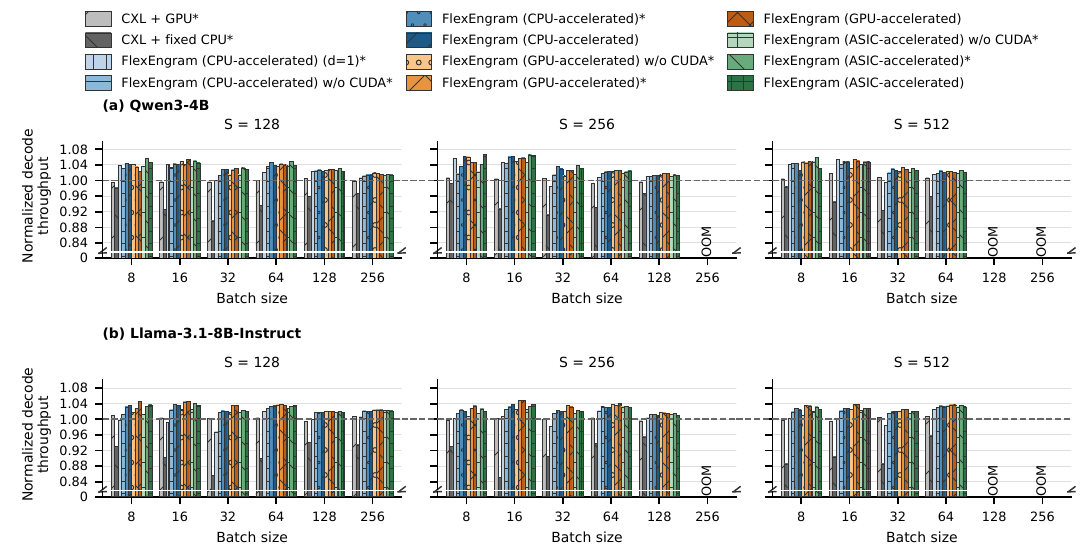}\par}%
\Description{Six grouped-bar panels compare ablations by batch size, using rows for Qwen and Llama and columns for sequence length. The dashed VanillaEngram reference is one. Colors and hatching identify execution paths and ablations; marked axis breaks preserve zero and OOM labels identify unavailable loads.}
\caption{Decode ablations on Qwen (a) and Llama (b), relative to VanillaEngram (1, dashed). OOM denotes unavailable loads; * denotes dual CXL-DRAM, and complete FlexEngram uses CXL-DRAM/CXL-PCM. w/o CUDA removes fusion.}
\label{fig:10}
\end{figure*}

\runinheading{Decode throughput} To isolate each mechanism, we add CPU-accelerated with fixed $d=1$, FlexScheduler, and CUDA fusion in sequence to CXL + fixed CPU, obtaining gains of 8.04\%, 8.33\%, 9.32\% on Qwen and 9.87\%, 12.05\%, 12.68\% on Llama (Fig.~\ref{fig:10}). GPU fusion raises Qwen/Llama gains from 8.96\%/12.29\% to 9.34\%/13.43\%. These gains reflect complementary mechanisms: acceleration shortens Gate--Conv, lookahead overlaps it with block computation, and fusion reduces GPU launches and memory accesses in Engram block 2. With these optimizations, replacing CXL-DRAM with higher-latency CXL-PCM for later Engram blocks' embeddings lowers mean throughput by only 0.11\% across both models and all three paths, indicating little end-to-end impact.

\subsection{Performance Comparison with Prior Art}

\begin{figure*}
{\centering
\includegraphics[width=\linewidth,alt={Six panels compare three prior methods with the three complete FlexEngram execution paths across two models, three sequence lengths, and batch sizes. Colors and hatching identify methods, dashed lines mark VanillaEngram, and axis breaks and OOM labels retain the plotting conventions.}]{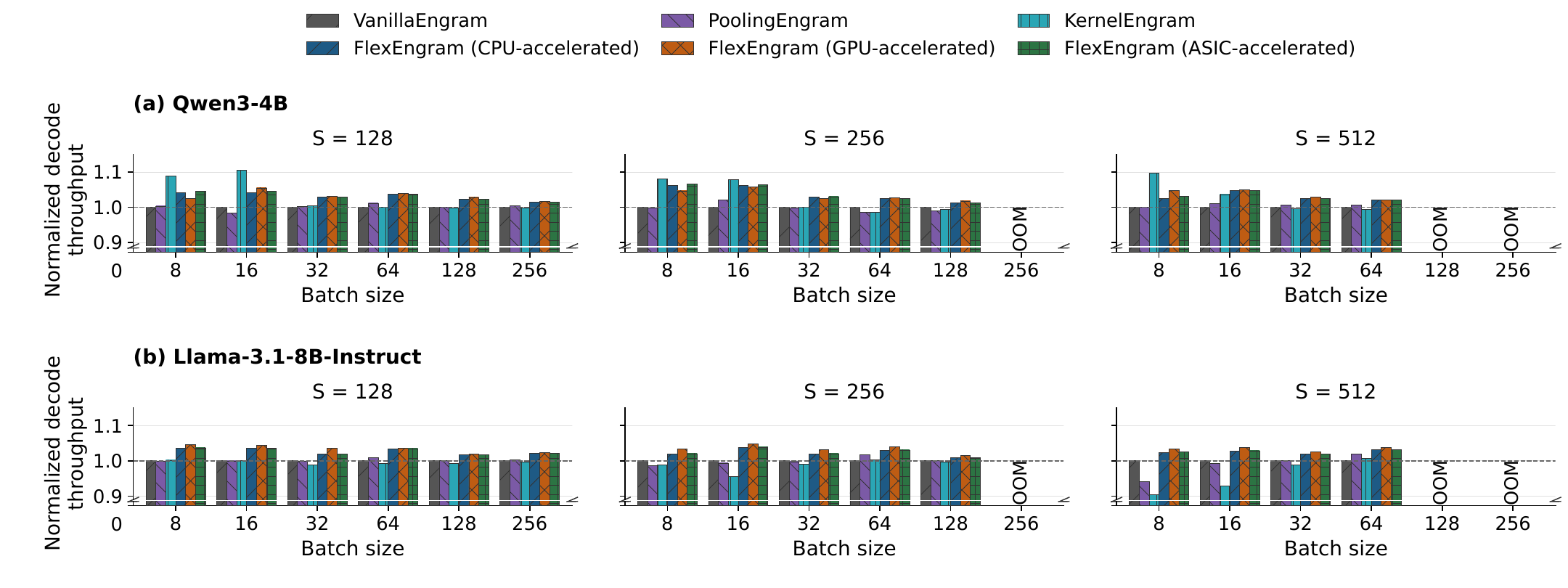}\par}%
\Description{Six panels compare three prior methods with the three complete FlexEngram execution paths across two models, three sequence lengths, and batch sizes. Colors and hatching identify methods, dashed lines mark VanillaEngram, and axis breaks and OOM labels retain the plotting conventions.}
\caption{Decode throughput comparison, normalized to VanillaEngram (1, dashed). OOM denotes unavailable loads.}
\label{fig:11}
\end{figure*}

\runinheading{Decode throughput} We compare complete FlexEngram with prior systems, obtaining mean gains across both models and the CPU-accelerated, GPU-accelerated, and ASIC-accelerated paths of 3.15\%, 3.23\%, and 2.67\% over VanillaEngram, PoolingEngram, and KernelEngram (Fig.~\ref{fig:11}). With only two Engram blocks, Engram occupies a small fraction of total LLM inference time, limiting end-to-end gains. KernelEngram's larger gains are concentrated in Qwen3-4B at small batch sizes, whereas all three FlexEngram paths improve throughput over both VanillaEngram and PoolingEngram across every evaluated non-OOM configuration on both models.

\subsection{Scaling to multiple Engrams}

\begin{figure*}
{\centering
\includegraphics[width=\linewidth,alt={Three panels scale from two to eight Engram blocks. Latency bars stack block 2 and the summed consumption latency of all later Engrams. Throughput uses No Engram as the reference, while accuracy uses residual proxies at three lookahead distances and a dashed original-model line. Shared legends identify execution paths and distances.}]{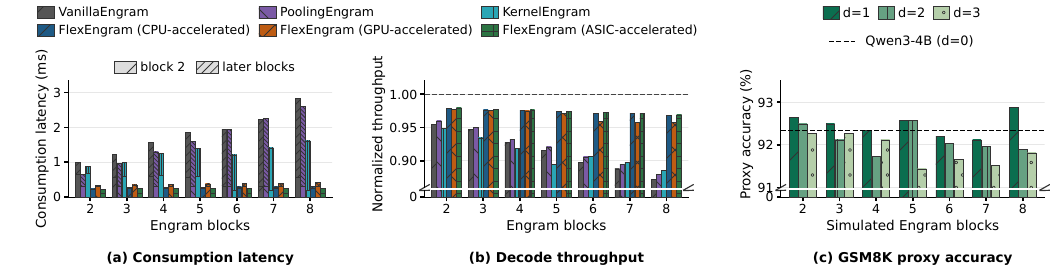}\par}%
\Description{Three panels scale from two to eight Engram blocks. Latency bars stack block 2 and the summed consumption latency of all later Engrams. Throughput uses No Engram as the reference, while accuracy uses residual proxies at three lookahead distances and a dashed original-model line. Shared legends identify execution paths and distances.}
\caption{Scaling Engram count on Qwen3-4B at block positions [2,15,18,21,24,27,30,35], B=64, S=512. (a) Stacked consumption latency of block 2 and all later Engrams. (b) Throughput normalized to No Engram (1, dashed). (c) GSM8K residual-proxy accuracy for $d=1,2,3$, with the original model dashed.}
\label{fig:12}
\end{figure*}

\runinheading{Multi-Engram consumption latency} Across seven layouts with two to eight Engrams, the mean aggregate consumption latency of later Engram blocks is \SI{0.030}{\milli\second} and \SI{0.052}{\milli\second} for CPU-accelerated and GPU-accelerated, respectively. Mean per-layout reductions relative to VanillaEngram are 97.57\% and 95.90\%, respectively (Fig.~\ref{fig:12}(a)). Including Engram block 2, total consumption averages \SI{0.279}{\milli\second}, \SI{0.375}{\milli\second}, and \SI{0.249}{\milli\second} for CPU-accelerated, GPU-accelerated, and ASIC-accelerated, respectively, versus \SI{1.803}{\milli\second} for VanillaEngram. Fusion, concurrency, and dedicated Engram accelerator pipelining overlap Gate--Conv with backbone work, limiting accumulated consumption.

\runinheading{Multi-Engram scaling overhead} To test scalability, we increase the number of Engrams from two to eight. With two-Engram throughput as the reference, total loss over these six additions is divided by six, giving a mean per-addition decline of 1.44\% for VanillaEngram versus 0.19\%, 0.35\%, and 0.18\% in the same path order (Fig.~\ref{fig:12}(b)). Averaging the three accelerated paths gives 0.24\%, an 83.46\% smaller decline. Losses with eight Engram blocks relative to No Engram are 3.20\%/4.27\%/3.10\% in that order, all below the 4.54\% loss of VanillaEngram with two Engram blocks. Overlap thus limits accumulated decode overhead as embedding capacity expands.

\runinheading{GSM8K accuracy} Across seven layouts, mean residual-proxy accuracy decreases overall from 92.46\% to 92.12\% and 91.87\% at $d=1,2,3$, versus the 92.34\% reference (Fig.~\ref{fig:12}(c)). At $d=1$, the mean residual-proxy accuracy is close to the original model, with no observed mean accuracy decrease. Within this proxy evaluation, this supports negligible mean accuracy loss for ASIC-accelerated with fixed one-block lookahead and CPU-accelerated configured with $d=1$. Even with eight simulated Engrams at $d=3$, proxy accuracy is 91.81\%, a loss of 0.53 percentage points. As a scale comparison across different models and evaluation protocols, this residual-proxy loss is substantially below Engram-27B's reported 2.2-percentage-point GSM8K 8-shot gain from two Engram blocks (2/15) \cite{cheng2026conditionalmemoryscalablelookup}. This small early-execution accuracy cost supports scaling to more Engram blocks and benefiting from their potential accuracy gains.

\subsection{Hardware Overhead}

To quantify dedicated Engram accelerator overhead, we synthesize Gate--Conv RTL with Yosys for 7-nm ASAP7 and use OpenSTA to analyze post-synthesis timing and estimate power from Qwen3-4B B8 decode SAIF activity. History SRAM uses abstract FakeRAM macros (Table~\ref{tab:3}).

\begin{table}[t]
\centering
\caption{Post-synthesis PPA per unit: ASAP7 RVT TT \cite{CLARK2016105}, \SI{0.7}{\volt}, \SI{25}{\degreeCelsius}, \SI{5.1}{\nano\second} (\SI{196.08}{\mega\hertz}).}
\label{tab:3}
\begin{tabular}{lrr}
\toprule
\textbf{Module} & \textbf{Power (\si{\milli\watt})} & \textbf{Area (\si{\milli\metre\squared})} \\
\midrule
Gate & 1073.13 & 4.054 \\
ShortConv & 967.14 & 4.553 \\
Control/stream & 553.69 & 2.323 \\
\textbf{Logic subtotal} & \textbf{2593.96} & \textbf{10.930} \\
History SRAM & 756.24 & 3.864 \\
\textbf{Total} & \textbf{3350.20} & \textbf{14.794} \\
\bottomrule
\end{tabular}
\end{table}

\runinheading{Area} The \SI{14.794}{\milli\metre\squared} unit comprises 58.18\% Gate--Conv, 26.12\% SRAM, and 15.70\% control/interfaces (Table~\ref{tab:3}). Thus, 84.30\% of its area serves Gate--Conv and history maintenance, keeping control/interface overhead limited. Its 64 contexts allocate \SI{10.97}{\mebi\byte} for \SI{6.75}{\mebi\byte} effective history. One later Engram block at B=256 requires four units (\SI{59.176}{\milli\metre\squared}); seven later Engram blocks at B=64 require 448 contexts/seven units (\SI{103.558}{\milli\metre\squared}). 

\begin{table}[t]
\centering
\caption{Component price assumptions (USD) \cite{10.1145/3805621.3807654,doi:10.1126/science.abj9979}.}
\label{tab:4}
\begin{tabular}{llr}
\toprule
\textbf{Component} & \textbf{Unit} & \textbf{Price} \\
\midrule
DRAM / CXL-DRAM & GB & 15.00 \\
CXL-PCM & GB & 9.00 \\
CXL switch & Switch & 5,800 \\
Host adapter & Adapter & 210 \\
Memory controller & Chip & 300 \\
\bottomrule
\end{tabular}
\end{table}

\runinheading{Timing margin} At \SI{196.08}{\mega\hertz} (\SI{5.1}{\nano\second}), worst setup and hold margins are 0.094 and \SI{0.020}{\nano\second}, with all electrical constraints satisfied. These positive margins show that the synthesized datapath meets the target clock under the evaluated conditions.

\runinheading{Power and core energy} The estimated per-unit core power is \SI{3.350}{\watt}, with 77.43\% logic/control and 22.57\% SRAM. At B=8, Qwen and Llama batch latencies are \SI{41.815}{\micro\second} and \SI{64.337}{\micro\second}; their per-token core energies are \SI{17.511}{\micro\joule} and \SI{26.943}{\micro\joule}, respectively. At the same power estimate, Llama's 1.54\(\times\) longer activity window explains its 1.54\(\times\) higher core energy. 

\subsection{Storage Economics}

\runinheading{Cost model} DeepSeek-V4.1-Flash's 200B FP8 embeddings occupy about \SI{200}{\giga\byte}; capacity adds one scale byte per 32 values. Using Table~\ref{tab:4} and PoolingEngram pricing \cite{10.1145/3805621.3807654}, we assume a historical PCM cost of 60\% of DRAM. Each Engram has about 100B embedding parameters; FlexEngram places the first Engram block's embeddings in CXL-DRAM and later Engram blocks' embeddings in CXL-PCM. Local DRAM replicates embeddings per node; a pool shares one copy/switch and adds an adapter/controller pair per node.

\begin{table}[t]
\centering
\caption{Pool savings over local DRAM (USD; local minus pool). Green/red: positive/negative; parentheses: Engram block counts.}
\label{tab:5}
\begin{tabular}{lrrrr}
\toprule
\textbf{Engram} & \textbf{Nodes} & \shortstack{\textbf{CXL-}\\\textbf{DRAM}} & \shortstack{\textbf{CXL-}\\\textbf{PCM}} & \textbf{FlexEngram} \\
\midrule
200B & 2 & \textcolor[RGB]{255,0,0}{-3,726.25} & \textcolor[RGB]{255,0,0}{-2,488.75} & \textcolor[RGB]{255,0,0}{-3,107.50} \\
(2) & 4 & \textcolor[RGB]{0,153,0}{1,441.25} & \textcolor[RGB]{0,153,0}{2,678.75} & \textcolor[RGB]{0,153,0}{2,060.00} \\
 & 8 & \textcolor[RGB]{0,153,0}{11,776.25} & \textcolor[RGB]{0,153,0}{13,013.75} & \textcolor[RGB]{0,153,0}{12,395.00} \\
 & 16 & \textcolor[RGB]{0,153,0}{32,446.25} & \textcolor[RGB]{0,153,0}{33,683.75} & \textcolor[RGB]{0,153,0}{33,065.00} \\
400B & 2 & \textcolor[RGB]{255,0,0}{-632.50} & \textcolor[RGB]{0,153,0}{1,842.50} & \textcolor[RGB]{0,153,0}{1,223.75} \\
(4) & 4 & \textcolor[RGB]{0,153,0}{10,722.50} & \textcolor[RGB]{0,153,0}{13,197.50} & \textcolor[RGB]{0,153,0}{12,578.75} \\
 & 8 & \textcolor[RGB]{0,153,0}{33,432.50} & \textcolor[RGB]{0,153,0}{35,907.50} & \textcolor[RGB]{0,153,0}{35,288.75} \\
 & 16 & \textcolor[RGB]{0,153,0}{78,852.50} & \textcolor[RGB]{0,153,0}{81,327.50} & \textcolor[RGB]{0,153,0}{80,708.75} \\
800B & 2 & \textcolor[RGB]{0,153,0}{5,555.00} & \textcolor[RGB]{0,153,0}{10,505.00} & \textcolor[RGB]{0,153,0}{9,886.25} \\
(8) & 4 & \textcolor[RGB]{0,153,0}{29,285.00} & \textcolor[RGB]{0,153,0}{34,235.00} & \textcolor[RGB]{0,153,0}{33,616.25} \\
 & 8 & \textcolor[RGB]{0,153,0}{76,745.00} & \textcolor[RGB]{0,153,0}{81,695.00} & \textcolor[RGB]{0,153,0}{81,076.25} \\
 & 16 & \textcolor[RGB]{0,153,0}{171,665.00} & \textcolor[RGB]{0,153,0}{176,615.00} & \textcolor[RGB]{0,153,0}{175,996.25} \\
\bottomrule
\end{tabular}
\end{table}

\begin{table*}
\centering
\caption{Engram attributes. Sharing spans instances; compute overlap uses earlier hidden states. FlexEngram predicts the minimum CPU-accelerated lookahead distance.}
\label{tab:6}
\begin{tabular}{l*{6}{c}}
\toprule
\textbf{Attribute} & \textbf{VanillaEngram} & \textbf{PoolingEngram} & \textbf{KernelEngram} & \textbf{PageGram} & \textbf{TF-Engram} & \textbf{FlexEngram} \\
\midrule
Reference & \cite{cheng2026conditionalmemoryscalablelookup} & \cite{10.1145/3805621.3807654} & \cite{du2026kernelsmithunifiedrecipeevolutionary} & \cite{lehmann2026pagegram} & \cite{ma2026tfengramtrainfreeengramssdbacked} & This work \\
Tiering by Engram block & \textcolor[RGB]{255,0,0}{\ding{55}} & \textcolor[RGB]{255,0,0}{\ding{55}} & \textcolor[RGB]{255,0,0}{\ding{55}} & \textcolor[RGB]{255,0,0}{\ding{55}} & \textcolor[RGB]{255,0,0}{\ding{55}} & \textbf{\textcolor[RGB]{0,153,0}{\ding{51}}} \\
Sharing of embeddings & \textcolor[RGB]{255,0,0}{\ding{55}} & \textcolor[RGB]{0,153,0}{\ding{51}} & \textcolor[RGB]{255,0,0}{\ding{55}} & \textcolor[RGB]{255,0,0}{\ding{55}} & \textcolor[RGB]{255,0,0}{\ding{55}} & \textbf{\textcolor[RGB]{0,153,0}{\ding{51}}} \\
Linear acceleration & \textcolor[RGB]{0,153,0}{\ding{51}} & \textcolor[RGB]{255,0,0}{\ding{55}} & \textcolor[RGB]{255,0,0}{\ding{55}} & \textcolor[RGB]{255,0,0}{\ding{55}} & \textcolor[RGB]{255,0,0}{\ding{55}} & \textbf{\textcolor[RGB]{0,153,0}{\ding{51}}} \\
Gate--Conv acceleration & \textcolor[RGB]{255,0,0}{\ding{55}} & \textcolor[RGB]{255,0,0}{\ding{55}} & \textcolor[RGB]{0,153,0}{\ding{51}} & \textcolor[RGB]{255,0,0}{\ding{55}} & \textcolor[RGB]{255,0,0}{\ding{55}} & \textbf{\textcolor[RGB]{0,153,0}{\ding{51}}} \\
Compute overlap & \textcolor[RGB]{255,0,0}{\ding{55}} & \textcolor[RGB]{255,0,0}{\ding{55}} & \textcolor[RGB]{255,0,0}{\ding{55}} & \textcolor[RGB]{255,0,0}{\ding{55}} & \textcolor[RGB]{255,0,0}{\ding{55}} & \textbf{\textcolor[RGB]{0,153,0}{\ding{51}}} \\
Adaptive scheduling & \textcolor[RGB]{255,0,0}{\ding{55}} & \textcolor[RGB]{255,0,0}{\ding{55}} & \textcolor[RGB]{255,0,0}{\ding{55}} & \textcolor[RGB]{255,0,0}{\ding{55}} & \textcolor[RGB]{0,153,0}{\ding{51}} & \textbf{\textcolor[RGB]{0,153,0}{\ding{51}}} \\
\bottomrule
\end{tabular}
\end{table*}

\runinheading{Storage cost} Savings relative to local DRAM increase with embedding parameter count and the number of sharing nodes (Table~\ref{tab:5}). At four to sixteen nodes, FlexEngram saves \$2,060--\$33,065 for 200B and \$33,616.25--\$175,996.25 for 800B. The latter's sixteen-node cost falls 88.89\%. Every tabulated FlexEngram configuration also costs less than CXL-DRAM-only, with tiering saving another \$618.75--\$4,331.25. 

\section{Related Work}\label{sec:VI}

\runinheading{Memory organization and access} Conditional-memory systems separate stored embeddings from sparse accesses. VanillaEngram retrieves entries selected by token IDs from host-DRAM embeddings \cite{cheng2026conditionalmemoryscalablelookup}; PoolingEngram shares CXL capacity across hosts and GPUs \cite{10.1145/3805621.3807654}. PageGram reduces flash reads through same-order n-gram head grouping within pages, NF4 storage, and caching \cite{lehmann2026pagegram}. TF-Engram tiers phrase memory across GPU, DRAM, and SSD and prefetches predicted next-token candidates \cite{ma2026tfengramtrainfreeengramssdbacked}.

FlexEngram selects a medium by Embedding access window for one Engram block's complete embeddings, rather than grouping vectors within a page. Early Engram blocks use low-latency CXL-DRAM; preceding Transformer computation covers CXL-PCM Embedding latency for the remaining Engram blocks.

Instances share static embeddings but retain private hidden states, key and value caches, Conv histories, and Embedding outputs.

In Table~\ref{tab:6}, a check mark denotes a mechanism explicitly proposed or described in the corresponding work; a cross denotes that the work does not propose it. These marks characterize each work's stated scope. The acceleration columns distinguish Linear from Gate--Conv, while compute overlap concerns hidden-state-dependent execution. This distinction separates reducing an operator's execution time from hiding that time behind backbone computation.

\runinheading{Engram computation} VanillaEngram combines multi-branch key/value Linear in one matrix multiplication \cite{cheng2026conditionalmemoryscalablelookup}, reducing calls. KernelEngram in Kernel-Smith uses execution feedback to search GPU implementations and fuses Gate, RMSNorm, depthwise Conv, and residual updates into two Triton kernels \cite{du2026kernelsmithunifiedrecipeevolutionary}. Linear fusion combines matrix work; Gate--Conv fusion reduces launches and intermediate accesses. FlexEngram combines both forms with heterogeneous overlap: AMX accelerates Linear, one CPU thread team fuses Gate--Conv, a separate GPU stream runs Gate--Conv concurrently, and channel-parallel engines of the dedicated Engram accelerator use banked history. All paths fuse common GPU operations in early Engram blocks.

\runinheading{Pipeline and scheduling} Engram, PoolingEngram, and PageGram overlap Embedding accesses determined by token IDs with Transformer computation \cite{cheng2026conditionalmemoryscalablelookup,10.1145/3805621.3807654,lehmann2026pagegram}; TF-Engram predicts future candidate accesses \cite{ma2026tfengramtrainfreeengramssdbacked}. Gate, however, depends on hidden states, so its execution remains on the backbone's critical path and is not hidden by these Embedding-overlap mechanisms. FlexEngram advances this dependent work with an earlier hidden state, preserving the backbone's in-order hidden-state computation. FlexScheduler predicts the minimum CPU-accelerated lookahead distance using computation, hidden-state input, delta return, and observed backbone timings; GPU-accelerated and ASIC-accelerated use fixed one-block lookahead. Adaptive scheduling means prefetch timing in TF-Engram and CPU-accelerated lookahead distance in FlexScheduler.

\section{Conclusion}\label{sec:VII}

FlexEngram shares embeddings and tiers them by Engram block, placing early Engram blocks' embeddings in CXL-DRAM and the rest in CXL-PCM. Only selected entries are delivered, while CPU-accelerated, GPU-accelerated, and ASIC-accelerated use earlier hidden states to overlap Gate--Conv with the backbone. The target adds delta to the backbone hidden state. FlexScheduler predicts the minimum CPU-accelerated lookahead distance, matching measured distances in 100\% of fifteen Llama workloads; GPU-accelerated and ASIC-accelerated use one-block lookahead. Across Qwen3-4B and Llama-3.1-8B-Instruct, average decode throughput improves by 3.15\% over VanillaEngram and 3.23\% over PoolingEngram. In Qwen's scaling experiment, mean loss per added Engram is 0.24\%, 83.46\% below VanillaEngram's decline; eight Engrams outperform VanillaEngram with two Engram blocks. The GSM8K residual proxy retains near-baseline mean accuracy at one-block lookahead, while larger lookahead distances increase mean accuracy loss, supporting the minimum distance needed to cover execution. These results support expanding conditional memory with minimal additional execution overhead.

\label{body:end}
\bibliographystyle{ACM-Reference-Format}
\bibliography{refs}

\end{document}